\documentclass[10pt,draft]{dis03}
\usepackage{epsf,amsmath}

\begin{document}

\title{HARD-SOFT POMERON TRANSITION FROM \\
QCD SATURATION IN THE DIPOLE PICTURE
}

\author{M.~LUBLINSKY \\
DESY, Theory \\
Hamburg 22607, Germany\\
E-mail: lublinm@mail.desy.de }

\maketitle

\begin{abstract}
\noindent We pose the question:
how far can we push the QCD dipole model and the idea of QCD saturation
to the kinematical regions  traditionally associated with soft 
nonperturbative physics? The answer is that it works sufficiently well
allowing a smooth transition from hard (BFKL) pomeron to the 
soft one. 
\end{abstract}

\begin{flushright}
\vspace{-7.5cm}
DESY-03-089\\
\today
\end{flushright}

\vspace{7.5cm}

\section{Introduction} 

We continue to develop a new global QCD analysis based on a solution of 
the Balitsky-Kovchegov (BK) non-linear evolution equation \cite{BK}. 
The general motivation rises from three main problems related to the DGLAP
evolution. First, it predicts a steep growth  of parton distributions
 at low $x$ which will eventually  violate  the unitarity constraints.
Second,  the twist OPE  breaks down at low $x$, when the higher twists 
become of the same order as the leading one. Finally,  the DGLAP
evolution is totally unable to describe low $Q^2$ data. Moreover, 
NLO corrections do not solve any of these problems.

The BK equation (BKE) is essentially the LO BFKL + 
unitarization (non-linearity).
We believe that this equation is a   solution to the above problems.
It accounts for the saturation effects  due to   high parton densities.
It sums high twist contributions and
 allows extrapolation to large distances.

In order to investigate the question formulated in the abstract
we adopt the following strategy. First, we 
construct a new saturation model ($\sigma_{dipole}$)
by solving the BKE and fitting free 
parameters to the low $x$ $F_2$ data. Then 
use thus obtained model beyond the limits of its formal applicability.

\section{Results}

In Ref. \cite{GLLM} we have constructed a saturation model based on a 
solution of the BKE. All available low $x$ data on the $F_2$ structure
function was successfully described. It includes the description at very low
photon virtuality down to $Q^2=0.045\,(GeV^2)\simeq \Lambda_{QCD}^2$.

One of the quantities associated with $F_2$ is $\lambda=d \ln F_2/d\ln (1/x)$.
In DIS $\lambda$ measures an effective pomeron intercept. 
For high $Q^2$,  $\lambda\simeq 0.3-0.4$ which is consistent with the data and 
 in agreement with the BFKL intercept. This is of no surprise
since the linear term in the BK equation is  the LO BFKL. Fig. 1
is our result for $\lambda$ at low $x$ and low $Q^2$. In this kinematics
$\lambda$ approaches the value $0.08-0.1$ which is the soft pomeron intercept.

Our saturation model provides a smooth interpolation between intercepts of 
the hard (BFKL) pomeron and the soft one.
\begin{figure}[!thb]
\vspace*{7.5cm}
\begin{center}
\includegraphics{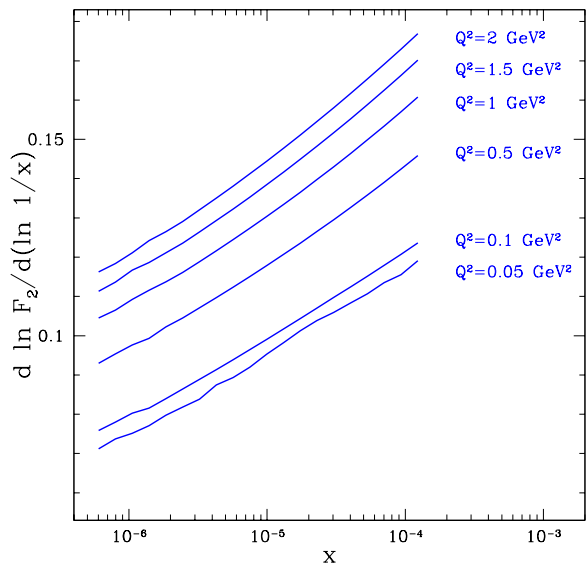}
\caption[*]{$\lambda$ at low $x$ and low $Q^2$.}
\end{center}
\label{lamda1}
\end{figure}
\vspace*{-0.5cm}

Being inspired by the success of the low $Q^2$ desription of the $F_2$
data we decided to push even further in the nonperturbative domain and
consider  total photoproduction \cite{BGLLM2}. To do so,
we need to introduce an additional nonperturbative parameter $Q_0$ which
relates the energy of the process $W$ to the Bjorken $x$: $x=Q_0^2/W^2$.
We find $Q_0^2=4\,m_q^2$ with the effective quark mass $m_q\simeq 0.15\,GeV$.
The two curves in Fig. \ref{lamda1} differ by the form of the secondary 
trajectory \footnote{The evolution of the secondary trajectories is
beyond the scope of our model.} added in order to describe the low energy data. 
\begin{figure}[!thb]
\vspace*{7.5cm}
\begin{center}
\includegraphics{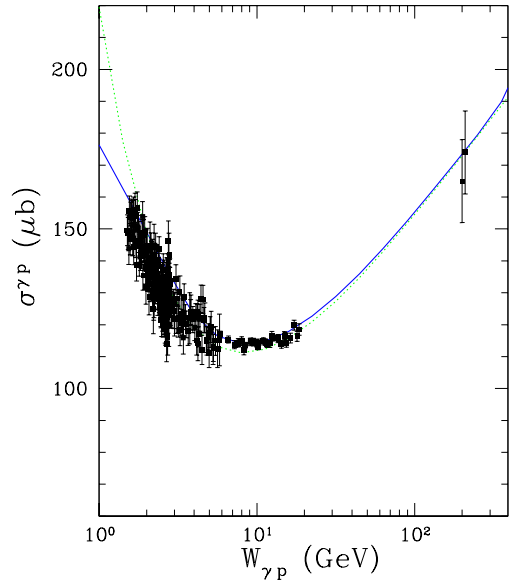}
\caption[*]{Total photoproduction.}
\end{center}
\end{figure}
\vspace*{-0.5cm}

Finally we would like to consider the hadron-hadron collision within
the same dipole approach \cite{BGLLM1}. 
We are pushing the dipole model deep into
the domain of nonperturbative QCD. We compute a hadron-proton cross section
as
$$
\sigma_{H-proton}(x) \; = \; \int d^2 r_\perp
 | \psi_{H}(r_\perp) |^2\,\sigma_{dipole}(r_\perp,x)
$$
We again introduce a nonperturbative scale $Q_0\,\propto \,M_H$ 
such that $x\,=\,Q_0^2/s$. $\sigma_{dipole}$ is the same dipole cross
section which was constructed to describe $F_2$ data. For the projectile 
hadron wave function $\psi_H$ a  Gaussian profile is used.

Figs. 3 and 4 present
 our results for meson-proton and proton-proton collisions. Except for 
$K$-meson a contribution of secondary reggeon $\propto s^{-0.45}$ was
added. We observe a reasonable agreement with the data.  

\begin{figure}[!thb]
\vspace*{7.5cm}
\begin{center}
\includegraphics{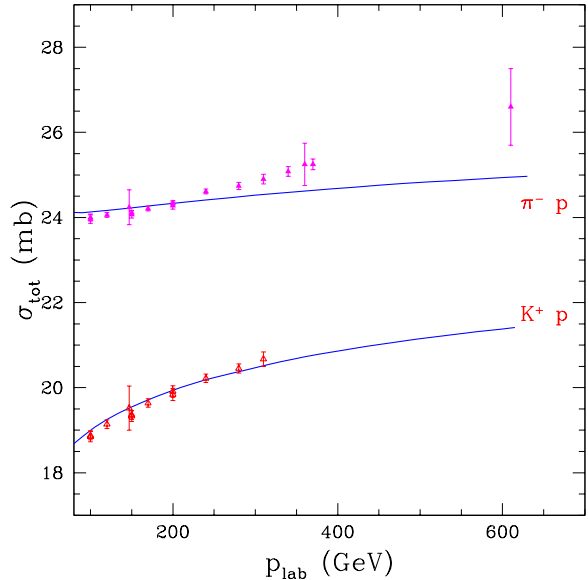}
\caption[*]{Meson-proton collision.}
\end{center}
\label{pip}
\end{figure}
\vspace*{-0.5cm}
\begin{figure}[!thb]
\vspace*{7.5cm}
\begin{center}
\includegraphics{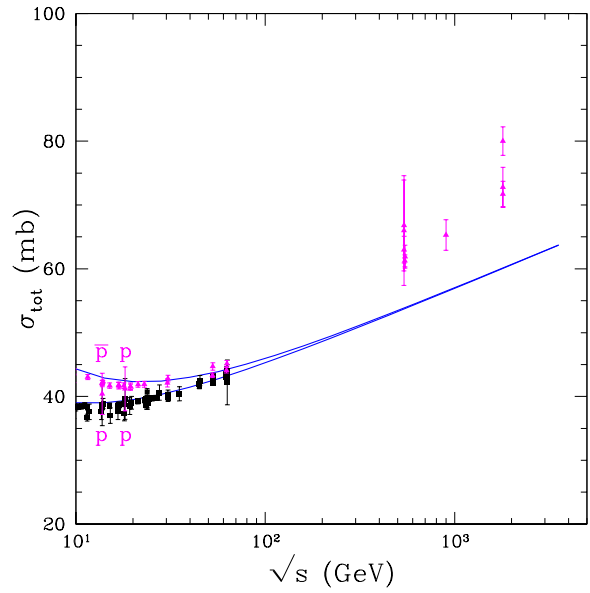}
\caption[*]{Proton-proton collision.}
\end{center}
\label{prot}
\end{figure}
\vspace*{-0.5cm}
\section{Summary}
$\bullet$ A new approach to global QCD analysis based on the BKE is developed.
$\bullet$  Low $x$ data on the $F_2$ structure function is
reproduced. 
$\bullet$ Our method allows extrapolation of the parton
distributions to the LHC energies and to low photon
virtualities $Q^2\ll 1\, GeV^2$.
$\bullet$ We find $\lambda \,\simeq\,0.25\,-\,0.4$ at large $Q^2$ (hard BFKL
pomeron) while
$\lambda\,\simeq\,0.08\, -\, 0.1$ at very low
$x$ and $Q^2$ well below $1 \,GeV^2$. A result which agrees with
the "soft pomeron" intercept without soft physics involved.
$\bullet$ The dipole picture pushed to a nonperturbative region works
       sufficiently well.  It describes  high energy photoproduction.
        Soft hadron interaction data are reproduced with quite satisfactory
         accuracy.

 Within the dipole picture  and QCD saturation, 
the soft pomeron is a 
phenomenon of multiple rescattering of the hard (BFKL) pomeron.

\section*{Acknowledgements} My gratitude is to my collaborators
J. Bartels, E. Gotsman, E. Levin,  U Maor.

\end{document}